\documentclass[fleqn,twoside,twocolumn,nofootinbib]{revtex4} 
\usepackage{ujp} 
\begin{document}
\title[POLAR PROPERTIES AND HYSTERESIS LOOPS]
{POLAR PROPERTIES  AND HYSTERESIS\\ LOOPS~\, IN~\, MULTILAYERED~\,
THIN~\, FILMS
FERROELECTRIC/VIRTUAL FERROELECTRIC}%
\author{E.A.~Eliseev}
\affiliation{V.E.~Lashkaryov Institute of Semiconductor Physics,
Nat.
Acad. of Sci. of Ukraine}
\address{41, Nauky Ave., Kyiv 03028, Ukraine}
\email{eugene.a.eliseev@gmail.com}
\affiliation{I.M.~Frantsevich Institute for Problems of Materials
Science,\\ Nat. Acad. of Sci. of Ukraine}%
\address{3, Krzhyzhanivskyi Str., Kyiv 03142, Ukraine}%
\author{M.D.~Glinchuk}%
\affiliation{I.M.~Frantsevich Institute for Problems of Materials
Science,\\ Nat. Acad. of Sci. of Ukraine}%
\address{3, Krzhyzhanivskyi Str., Kyiv 03142, Ukraine}%
\author{A.N.~Morozovska}
\affiliation{V.E.~Lashkaryov Institute of Semiconductor Physics,
Nat.
Acad. of Sci. of Ukraine}%
\address{41, Nauky Ave., Kyiv 03028, Ukraine}%
\email{eugene.a.eliseev@gmail.com}%
\author{Ya.V.~Yakovenko}%
\affiliation{Taras Shevchenko National University of Kyiv, Faculty of Physics}%
\address{2, Academician Glushkov Ave., Bld.~1, Kyiv 03127, Ukraine}%
\udk{}  \razd{\secix}

\setcounter{page}{1038}%
\maketitle

\begin{abstract}
In the framework of Landau--Ginzburg--Devonshire (LGD)
phenomenological theory, the influence of misfit strains, surface
energy, and finite-size effects on phase diagrams, polar properties,
and hysteresis loops has been calculated for multilayered thin films
of the type ferroelectric/virtual ferroelectric. The influence of
elastic deformations that arise at the interface thin
film--substrate owing to a mismatch between the lattice constants in
the film and the substrate on the phase diagrams of multilayered
thin films virtual ferroelectric SrTiO$_{3}$/ferroelectric
BaTiO$_{3}$ has been studied for the first time. In contrast to bulk
BaTiO$_{3}$, in which only four phases (cubic, tetragonal,
orthorhombic, and rhombohedral) can exist, it turned out that six
thermodynamically stable BaTiO$_{3}$ phases (paraelectric phase and
tetragonal (FEc), two monoclinic (FEaac and FEac), and two
orthorhombic (FEa and FEaa) ferroelectric phases) can exist in
multilayered SrTiO$_{3}$/BaTiO$_{3}$ films. The main polar
properties of hysteresis loops (shape, coercive field, and
spontaneous polarization) in thin multilayered
SrTiO$_{3}$/BaTiO$_{3}$ films are calculated. It is shown that the
system demonstrates a strong dependence of its polar properties on
the thickness of SrTiO$_{3}$ and BaTiO$_{3}$ layers, as well as on
the elastic misfit strains, with SrTiO$_{3}$ playing the role of
dielectric layer: the thicker the layer, the stronger is the
depolarization field, which, in its turn, reduces the spontaneous
polarization in the BaTiO$_{3}$ film.
\end{abstract}

\section{Introduction}

Multilayer ferroelectric films are challenging objects for experimental and
theoretical researches, because they demonstrate enhanced polar properties
in comparison with monolayer films of the same thickness. However, the
influence of misfit strains that arise owing to the difference between the
elastic properties of the substrate and the multilayer film on the phase
diagrams and the polar properties of and the hysteresis loops in multilayer
thin films of the type ferroelectric/virtual ferroelectric has almost not
been studied theoretically.

\subsection{Current state of experimental researches dealing with the polar
properties of multilayer ferroelectric films}

Multilayer ferroelectric films and compositionally graded heterostructures
demonstrate an enhanced spontaneous polarization, a huge pyroelectric
response, and a large piezoelectric deformation (see, e.g., work
\cite{Nath:2008} and the references therein). Electric and electromechanical
properties of complicated oxide superfilms CaTiO$_{3}$/BaTiO$_{3}$ are
closely connected with the structure evolution and the electric polarization
in every layer induced by an applied electric field.  The coexistence of
elastic deformations and rotations of oxygen octahedra near the interface
between polar and nonpolar film layers allows this system to show a rich
spectrum of nanoscaled phenomena. Multilayer materials with improved
functional characteristics are of interest for the development of
nanophysics and nanotechnology. For instance, a piezoelectric response to an
applied electric field arises in the CaTiO$_{3}$ dielectric layers of
nano-sized multilayer dielectric-ferroelectric films CaTiO$_{3}$/BaTiO$_{3}$
\cite{Young:2010,Young:2011}. The measured piezoelectric coefficient of
54~pm/V agrees with the results of \textit{ab initio} calculations, in which
the tetragonal symmetry in the superfilm is supposed to be induced by the
SrTiO$_{3}$ substrate.

Stephanovich {\it et al.} \cite{Stephanovich:2005} studied the
ferroelectric phase transition and the formation of domains in a
periodic multilayer film consisting of alternating ferro- and
paraelectric nano-sized layers. They found that ferroelectric
domains formed in different ferroelectric layers can interact with
one another through the paraelectric layer. By minimizing the
Ginzburg--Landau functional, those authors calculated the critical
transition temperature $T_{c}$ as a function of the thickness of
a multilayer \mbox{ferro-/paraelectric} film. However, the influence
of the strains arising owing to a misfit between the substrate and
the multilayer film on the polar properties of the system was not
studied.

The theoretical approach is considered to be necessary in the effective search
for thin films and their multilayer structures with considerably improved
functional properties \cite{Schlom:2008}. The development and the application of
methods aimed at growing the thin films of functional oxides are at the
beginning, and plenty of obstacles should be overcome expectedly for a
better structural monitoring of layer quantity to be developed, as well as
for the reproducibility and the monitoring of the corresponding
electrophysical, polar, optical, and spintronic properties to be achieved.

\subsection{Current state of theoretical researches dealing with the polar
properties of multilayer ferroelectric films}

\subsubsection*{1.2.1. Theoretical calculations of polar
properties\\ \phantom{1.2.1. }of multilayer ferroelectric films}

\textit{Ab initio} calculations \cite{Xifan:2011} of polar and piezoelectric
properties of a CaTiO$_{3}/$BaTiO$_{3}$ superfilm grown up on a SrTiO$_{3}$
substrate showed that short-period superfilms have a higher spontaneous
polarization than that of long-period ones. A thermodynamic model
\cite{Roytburd:2005} for the polarization and the dielectric
response of ferro-/paraelectric bilayers and multilayer films was proposed.
The strong electrostatic coupling between the layers was shown to suppress the
ferroelectric behavior at a critical thickness of the paraelectric layer. A
bilayer is characterized by a giant dielectric response, which is similar to
the dielectric anomaly in a vicinity of the Curie--Weiss temperature in a
homogeneous ferroelectric with a critical thickness. The numerical analysis
carried out for a pseudomorphic (001) heteroepitaxial BaTiO$_{3}$/SrTiO$_{3}$
bilayer on a (001) SrTiO$_{3}$ substrate and an undeformed
BaTiO$_{3}$/SrTiO$_{3}$ bilayer showed that the polarization and the dielectric peak
disappear totally at SrTiO$_{3}$ contents of about 66 and 14\% in those two systems,
respectively.

It was shown from the first principles that the ground state of a
system consisting of a multilayer film of the type
ferroelectric/paraelectric, PbTiO$_{3}$/SrTiO$_{3}$, is not purely
ferroelectric. It also includes antiferrodistortive rotations of
oxygen octahedra, which are an analog of nonintrinsic
ferroelectricity. In addition, the dielectric response of the system
was found to have a weak temperature dependence. The results
obtained theoretically were demonstrated to agree with those of
experimental observations \cite{Bousquet:2008}.

Fong {\it et al.} \cite{Fong:2004} connected the critical character
of the thickness of a thin ferroelectric film with the presence of a
near-surface layer, which includes rotated oxygen octahedra
reconstructed into a surface antiferrodistortive structure.

\subsubsection*{1.2.2. Theoretical calculations of polar properties of\\ \phantom{1.2.1. }multilayer
ferroelectric films in the framework\\ \phantom{1.2.1. }of LGD
phenomenological theory}

Numerical calculations were carried out for heteroepitaxial (001) bilayers
and (001) multilayer PbTiO$_{3}$/SrTiO$_{3}$ (PTO/STO) structures on (001)
STO substrates. Similarly to the temperature dependence inherent to a
homogeneous ferroelectric, there exists a critical value for the PTO layer,
below which the bilayer or multilayer film is in the paraelectric state.
This critical thickness depends strongly on the density of coupled charges
at the interface between the films. The results of calculations for the
dielectric response show that the ferroelectric phase transition, which is
well-pronounced for uncharged bilayers and superfilms, gets smeared, and the
total dielectric constant decreases, as the charge density increases
\cite{Okatan:2010,Okatan:1}. In works \cite{Okatan:2010,Okatan:1}, the nonlinear
thermodynamic model was used to study the role of interfaces in the
polarization response of a ferroelectric-paraelectric bilayer, and the
numerical analysis was made for a BTO/STO bilayer. The critical thickness was
predicted for SrTiO$_{3}$, at which the spontaneous polarization in
BaTiO$_{3}$ should disappear owing to the growth of the depolarization field. The
critical thickness was demonstrated to decrease together with a reduction of
the total bilayer thickness, by indicating the fact that the interface effects are
more pronounced in thinner bilayers \cite{Misirlioglu:2007}.

The LGD formalism enables the ferroelectric properties of many
materials to be described adequately. In most cases, only one phase
characterized by the components of its polarization oriented in the
film plane was revealed in thin films of ferroelectric perovskites
(see, e.g., works
\cite{Pertsev:2003,Sharma:2004,Stephanovich:2006}). In particular,
the properties of phase transitions and mechanisms of domain
structure formation in films consisting of alternating ferroelectric
and paraelectric nano-sized layers were studied. The domain
structure period and the temperatures of phase transitions were
found to depend on the thickness of ferro- and paraelectric layers
\cite{Stephanovich:2005}.

By analyzing the literature, it is possible to conclude that, up to now, the influence of
misfit strains, which arise owing to the difference between the elastic
properties of the substrate and the multilayer film, on the phase diagrams,
polar properties, and hysteresis loops in multilayer thin films of the type
ferroelectric/virtual ferroelectric has not been studied enough
theoretically. Therefore, in what follows, we consider multilayer films
consisting of films of different materials.

\section{Mathematical Formulation of the Problem}

This work is aimed at (i)~calculating the influence of misfit strains, surface
energy, and size effects (i.e. the dependence on the ratio between the layer
thicknesses) on the phase diagrams, polar properties, and hysteresis loops
in thin multilayer films of the type ferroelectric/virtual ferroelectric
material and (ii)~establishing the influence of elastic deformations that
arise at the interface between the thin film and the substrate owing to a
mismatch between their lattice constants on the phase diagrams of thin
multilayer thin films virtual ferroelectric (SrTiO$_{3})$/ferroelectric
(BaTiO$_{3}$). Probable monodomain configurations of the polarization in
a multilayer BTO/STO film are illustrated in Fig.~1.

\begin{figure}
\includegraphics[width=\column]{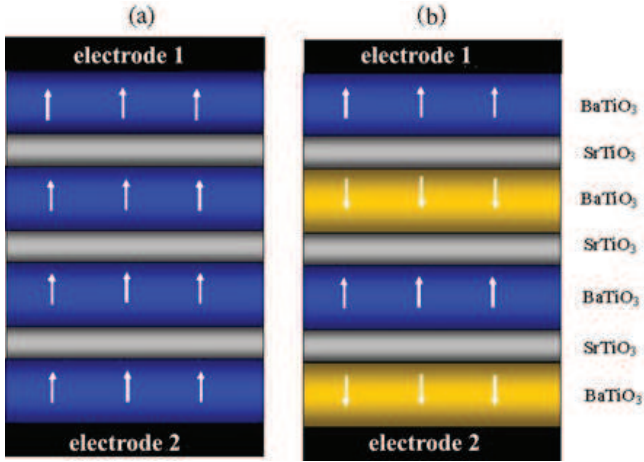}
\vskip-3mm\caption{Probable polarization configurations in
a multilayered BTO/STO film  }
\end{figure}

Monodomain states of the system can be described analytically,
because the corresponding problem is reduced to a one-dimensional
one. Those states can be realized in the case of defect-free
multilayer films sandwiched between two conducting electrodes. The
monodomain state has a lower correlation energy and a lower energy
of domain walls than the polydomain one has. For a film without
electrodes (or with semiconductor electrodes), the polydomain state
is energetically beneficial, because it has a lower depolarization
field, although the energy of domain walls is higher. As is shown
below, the monodomain state corresponds expectedly to solitary
hysteresis loops of ferroelectric polarization.\looseness=1

The free energy functional for the multilayer system depicted in
Fig.~1 includes the energy of every layer and the energies of
interfaces,
\begin{subequations}
\begin{equation}
F=F_{\rm layer}+F_{\rm interface},   \label{eq1}
\end{equation}
\begin{equation}
F_{\rm
layer}=\sum\limits_{i=1}^{n}{\int\limits_{z_{i-1}}^{z_{i}}{\Delta
F_{b}^{(i)}\,\left( {P_{1}^{(i)},P_{2}^{(i)},P_{3}^{(i)}}\right)
dz}},
 \label{eq2}
\end{equation}
\begin{widetext}
\begin{equation}
F_{\rm interface}=\left( {\begin{array}{c}
\sum\limits_{i=1}^{n}{\left( {\begin{array}{l} \left. {\left(
{\frac{\alpha _{l1}^{(i)}}{2}\left( {P_{1}^{(i)}}\right)
^{2}+\frac{\alpha _{l2}^{(i)}}{2}\left( {P_{2}^{(i)}}\right)
^{2}+\frac{\alpha _{l3}^{(i)}}{2}\left( {P_{3}^{(i)}}\right)
^{2}}\right) }\right\vert
_{z=z_{i-1}}+ \\
+\left. {\left( {\frac{\alpha _{r1}^{(i)}}{2}\left( {P_{1}^{(i)}}\right)
^{2}+\frac{\alpha _{r2}^{(i)}}{2}\left( {P_{2}^{(i)}}\right)
^{2}+\frac{\alpha _{r3}^{(i)}}{2}\left( {P_{3}^{(i)}}\right) ^{2}}\right) }\right\vert
_{z=z_{i}} \\ \end{array}}\right) +} \\
+\sum\limits_{i=1}^{n-1}{\left. {\left( {\frac{\gamma
_{1}^{(i)}}{2}\left( {P_{1}^{(i)}-P_{1}^{(i+1)}}\right)
^{2}+\frac{\gamma _{2}^{(i)}}{2}\left(
{P_{2}^{(i)}-P_{2}^{(i+1)}}\right) ^{2}+\frac{\gamma
_{3}^{(i)}}{2}\left(
{P_{3}^{(i)}-P_{3}^{(i+1)}}\right) ^{2}}\right) }\right\vert _{z=z_{i}}} \\
\end{array}}\right) .   \label{eq3}
\end{equation}%
\end{widetext}
\end{subequations}
Here, we introduced the following notation: $h_{i}$ is the
thickness of the $i$-th layer, and $z_{i}$ is the position of
interface between different layers, so that
$z_{i}=\sum\nolimits_{j=1}^{j=i}{h_{j}}$ and $z_{0}=0$. The surface
energy (\ref{eq3}) is similar to that of a monolayer.
The last term corresponds to the interaction between polarization
components belonging to neighbor layers \cite{Zhong:1997}. The
expression for the surface energy includes only the terms, which are
allowed by the tetragonal symmetry. The terms linear in
$P_{3}^{(i)}$ are excluded, because they would result in a
spontaneous polarization of the nonpolar paraelectric phase. From the
condition of polarization continuity, we adopt that $\alpha
_{rj}^{(i)}=\alpha _{lj}^{(i+1)}=\gamma _{j}^{(i)}$ in
Eq.~(\ref{eq3}). The surface terms in Eq.~(\ref{eq3}) have a meaning
of the squared polarization gradient between layers (see work
\cite{Chew:2007}); in particular, in the limiting case $\gamma
_{j}^{(i)}\rightarrow \infty $, the polarization on the surface
$P_{j}^{(i)}\equiv P_{j}^{(i+1)}$.

The density of the bulk part of the free energy $\Delta F_{b}$ for
the paraelectric phase with initial symmetry $m3m$, which
corresponds to the symmetry of every BaTiO$_{3}$ or SrTiO$_{3}$
layer, can be written down as a Taylor expansion in the polarization
components $P_{k}$ $(k=1,2,3)$:\looseness=1
\begin{widetext}\vskip-5mm
\begin{equation}
\Delta F_{b}^{(i)}\left( {P_{1},P_{2},P_{3}}\right) \,=\left(
{\begin{array}{l}
\frac{a_{1}^{(i)}}{2}\left( {P_{1}^{2}+P_{2}^{2}}\right)
+\frac{a_{3}^{(i)}}{2}P_{3}^{2}+\frac{a_{12}^{(i)}}{2}P_{1}^{2}P_{2}^{2}+\frac{a_{13}^{(i)}}{2}%
P_{3}^{2}\left( {P_{2}^{2}+P_{1}^{2}}\right) + \\[2mm]
+\frac{a_{11}^{(i)}}{4}\left( {P_{1}^{4}+P_{2}^{4}}\right) +\frac
{a_{33}^{(i)}}{4}P_{3}^{4}+\frac{a_{111}^{(i)}}{6}\left(
{P_{1}^{6}+P_{2}^{6}+P_{3}^{6}}\right)
+\frac{a_{123}^{(i)}}{2}P_{3}^{2}P_{2}^{2}P_{1}^{2}+ \\[2mm]
+\frac{a_{112}^{(i)}}{4}\left( {P_{3}^{4}\left(
{P_{1}^{2}+P_{2}^{2}}\right) +P_{1}^{4}\left(
{P_{2}^{2}+P_{3}^{2}}\right) +P_{2}^{4}\left(
{P_{1}^{2}+P_{3}^{2}}\right) }\right) + \\[2mm]
+\frac{g_{11}^{(i)}}{2}\left(
{\frac{\partial{\kern1pt}P_{3}}{\partial {\kern1pt}z}}\right)
^{2}+\frac{g_{44}^{(i)}}{2}\left( {\left(
{\frac{\partial{\kern1pt}P_{1}}{\partial{\kern1pt}z}}\right)
^{2}+\left(
{\frac{\partial{\kern1pt}P_{2}}{\partial{\kern1pt}z}}\right)
^{2}}\right) -\left.
{\frac{1}{2}P_{3}E_{3}^{d(i)}-P_{3}E_{3}^{e}}\right) \\ \end{array}
}\right).   \label{eq4}
\end{equation}
\end{widetext}%

\noindent Here, we used the Voigt notation and the Cartesian
coordinates. The surface cross-section (100) is considered. The
quantities $g_{ij}$ are the gradient coefficients of the
ferroelectric. We also take into account that the equilibrium state
for the thin film on the substrate will be found as a minimum of the
Helmholtz free energy $F=G_{+}\int_{V}u_{jk}X_{jk}{d^{3}r}$, which
can be obtained from the Hibbs free energy using the Legendre
transformation \cite{Pertsev:1998}.

The mismatch of the lattice constants between the
film and the substrate causes a redefinition of the free energy
coefficients due to the misfit strain $u_{m}^{(i)}=\frac{c_{i}-c_{S}}{c_{S}}$
that emerges between the multilayer film and the substrate,
\begin{subequations}
\[ a_1^{(i)} =2\alpha _1^{(i)} \left( T \right)-\frac{2\left(
{Q_{11}^{(i)} +Q_{12}^{(i)} } \right)\,}{s_{11}^{(i)} +s_{12}^{(i)}
}u_m^{(i)} ,\]
\begin{equation}
\label{eq5} a_3^{(i)} =2\alpha _1^{(i)} \left( T
\right)-\frac{4Q_{12}^{(i)} \,}{s_{11}^{(i)} +s_{12}^{(i)}
}u_m^{(i)} ,
\end{equation}
\[
a_{11}^{(i)} =4\alpha _{11}^{(i)} +2\frac{\left( {Q_{11}^{(i)2}
+Q_{12}^{(i)2} } \right)s_{11}^{(i)} -2s_{12}^{(i)} Q_{11}^{(i)}
Q_{12}^{(i)} }{s_{11}^{(i)2} -s_{12}^{(i)2} },\]
\begin{equation}
\label{eq6}
 a_{33}^{(i)}
=4\alpha _{11}^{(i)} +\frac{4Q_{12}^{(i)2} }{s_{11}^{(i)}
+s_{12}^{(i)} },
\end{equation}
\[
a_{12}^{(i)}\! =\!2\alpha _{12}^{(i)} \!-\!2\frac{\left(
{Q_{11}^{(i)2} \!+\!Q_{12}^{(i)2} } \right)s_{12}^{(i)}
-2s_{11}^{(i)} Q_{11}^{(i)} Q_{12}^{(i)} }{s_{11}^{(i)2}
\!-\!s_{12}^{(i)2} }\!+\!\frac{Q_{44}^{(i)2} }{s_{44}^{(i)} },\]
\begin{equation}
\label{eq7}
 a_{13}^{(i)} =2\alpha _{13}^{(i)}
+\frac{2Q_{12}^{(i)} \left( {Q_{11}^{(i)} +Q_{12}^{(i)} }
\right)\,}{s_{11}^{(i)} +s_{12}^{(i)} }.
\end{equation}
\end{subequations}
Here, $\alpha _{i}^{(i)}$ and $\alpha _{ij}^{(i)}$ are the
dielectric rigidity and the rigidity coefficients of higher order at
a constant pressure, $s_{11}$ and $s_{12}$ are the coefficients of
elastic compliance, $Q_{ij}$ is the electrostriction tensor, and
$u_{m}$ is the misfit strain between the film and the substrate. The
influence of misfit dislocations stimulates a relaxation of misfit
strains. This fact can be taken into account by redefining $u_{m}$
(see, e.g., \cite{Speck:1994}).

The quantity $E_{3}^{e}$ in Eq.~(\ref{eq4}) is the external field,
and $E_{3}^{d}$ is the depolarization field arising owing to an
incomplete screening of the polarization by the environment, a
non-uniform distribution of the polarization, and/or its disappearance at
the surface. For a one-dimensional distribution, the quantity
$E_{3}^{d}$ has the form
\begin{equation}
E_{3}^{(i)}=\frac{1}{\varepsilon_{0}\varepsilon_{b}^{(i)}}\left(
{\frac {\sum\limits_{i=1}^{n}{{h_{i}}\mathord{\left/ {\vphantom
{{h_i } {\varepsilon _b^{(i)} }}} \right. \kern-\nulldelimiterspace}
{\varepsilon _b^{(i)} }}\left\langle {P_{3}^{(i)}}\right\rangle
}{\sum\limits_{i=1}^{n}{{h_{i}}\mathord{\left/ {\vphantom {{h_i }
{\varepsilon _b^{(i)} }}} \right. \kern-\nulldelimiterspace}
{\varepsilon _b^{(i)} }}}-P_{3}^{(i)}}\right), \label{eq8}
\end{equation}
where $\varepsilon_{b}^{(i)}$ is the background dielectric
permittivity of the $i$-th layer, which is not connected with the
soft optical mode \cite{Rupprecht:1964}, and $\left\langle
{P_{3}^{(i)}}\right\rangle =\frac
{1}{h_{i}}\int_{z_{i}}^{z_{i}+h_{i}}{P_{3}^{(i)}dz}$ is the average
polarization of the $i$-th layer.

Below, we consider a periodic multilayer film with the geometrical period
consisting of two layers, $h_{1}$ and $h_{2}$. The polarization distribution
over different superperiods $h_{1}+h_{2}$ is supposed to be the same,
which is valid, if the coefficients in the surface energy are independent of
the surface arrangement. In the case of a one-dimensional distribution, the
expression for the depolarization field can be simplified as
follows:
\[
E_3^{(1)} =\frac{1}{\varepsilon _0 \varepsilon _b^{(1)} }\times
\]
\begin{equation}
\label{eq9} \times\left( {\frac{{\left\langle {P_3^{(1)} }
\right\rangle h_1 } \mathord{\left/ {\vphantom {{\left\langle
{P_3^{(1)} } \right\rangle h_1 } {\varepsilon _b^{(1)} }}} \right.
\kern-\nulldelimiterspace} {\varepsilon _b^{(1)} }\!+\!{\left\langle
{P_3^{(\ref{eq4})} } \right\rangle h_2 } \mathord{\left/ {\vphantom
{{\left\langle {P_3^{(\ref{eq4})} } \right\rangle h_2 } {\varepsilon
_b^{(\ref{eq4})} }}} \right. \kern-\nulldelimiterspace} {\varepsilon
_b^{(\ref{eq4})} }}{{h_1 } \mathord{\left/ {\vphantom {{h_1 }
{\varepsilon _b^{(1)} }}} \right. \kern-\nulldelimiterspace}
{\varepsilon _b^{(1)} }\!+\!{h_2 } \mathord{\left/ {\vphantom {{h_2
} {\varepsilon _b^{(\ref{eq4})} }}} \right.
\kern-\nulldelimiterspace} {\varepsilon _b^{(\ref{eq4})}
}}\!-\!P_3^{(1)} } \right),
\end{equation}
\[
E_3^{(\ref{eq4})} \!=\!\frac{1}{\varepsilon _0 \varepsilon
_b^{(\ref{eq4})} }\times
\]
\begin{equation}
\label{eq10} \times\left( {\frac{{\left\langle {P_3^{(1)} }
\right\rangle h_1 } \mathord{\left/ {\vphantom {{\left\langle
{P_3^{(1)} } \right\rangle h_1 } {\varepsilon _b^{(1)} }}} \right.
\kern-\nulldelimiterspace} {\varepsilon _b^{(1)} }\!+\!{\left\langle
{P_3^{(\ref{eq4})} } \right\rangle h_2 } \mathord{\left/ {\vphantom
{{\left\langle {P_3^{(\ref{eq4})} } \right\rangle h_2 } {\varepsilon
_b^{(\ref{eq4})} }}} \right. \kern-\nulldelimiterspace} {\varepsilon
_b^{(\ref{eq4})} }}{{h_1 } \mathord{\left/ {\vphantom {{h_1 }
{\varepsilon _b^{(1)} }}} \right. \kern-\nulldelimiterspace}
{\varepsilon _b^{(1)} }\!+\!{h_2 } \mathord{\left/ {\vphantom {{h_2
} {\varepsilon _b^{(\ref{eq4})} }}} \right.
\kern-\nulldelimiterspace} {\varepsilon _b^{(\ref{eq4})}
}}\!-\!P_3^{(\ref{eq4})} } \right).
\end{equation}%
One can see that the normal component of the polarization in a layer affects the
normal component of the polarization in the other layer by means of the internal
electric field.

\section{Coupled Equations}

The variation of the free energy functional brings about the following
system of nonlinear Euler--Lagrange equations with the corresponding
boundary conditions:
\begin{subequations}
\begin{equation}
\label{eq11} \left\{ {\begin{array}{l} ( a_3^{(i)} +a_{13}^{(i)} (
P_2^{(i)}  )^2+a_{13}^{(i)}
( P_1^{(i)}  )^2 )P_3^{(i)} +\\[2mm]
+a_{33}^{(i)} ( P_3^{(i)}  )^3+f_3^{(i)} -g_{11}^{(i)}
\frac{\partial ^2}{\partial
\,z^2}P_3^{(i)} =E_3^{(i)} \,, \\[2mm]
  \Biggl( \alpha _{S3}^{(i)} P_3^{(i)} +\gamma _3^{(i-1)} (
P_3^{(i)} -P_3^{(i-1)}  )-\\
-g_{11}^{(i)} \frac{dP_3^{(i)} }{dz}
\Biggr) \Biggr|_{z=z_{i-1} } =0, \\
  \Biggl( \alpha _{S3}^{(i)} P_3^{(i)} +\gamma _3^{(i)} (
{P_3^{(i)} -P_3^{(i+1)} } )+\\
+g_{11}^{(i)} \frac{dP_3^{(i)} }{dz}
\Biggr) \Biggr|_{z=z_i } =0 \,.  \\
 \end{array}} \right.
\end{equation}
\begin{equation}
\label{eq12} \left\{ {\begin{array}{l} ( {a_1^{(i)} +a_{12}^{(i)} (
{P_1^{(i)} } )^2+a_{13}^{(i)}
( {P_3^{(i)} } )^2} )P_2^{(i)} +\\ \\[1mm]
+a_{11}^{(i)} ( {P_2^{(i)} } )^3+f_2^{(i)} -g_{44}^{(i)}
\frac{\partial ^2}{\partial
\,z^2}P_2^{(i)} =0\,, \\ \\[1mm]
\Biggl( \alpha _{S2}^{(i)} P_2^{(i)} +\gamma _2^{(i-1)} (
{P_2^{(i)} -P_2^{(i-1)} } )-\\
-g_{44}^{(i)} \frac{dP_2^{(i)} }{dz}
\Biggr) \Biggr|_{z=z_{i-1} } =0, \\
  \Biggl( \alpha _{S2}^{(i)} P_2^{(i)} +\gamma _2^{(i)} (
{P_2^{(i)} -P_2^{(i+1)} } )+\\
+g_{44}^{(i)} \frac{dP_2^{(i)} }{dz}
\Biggr) \Biggr|_{z=z_i } =0\,. \\
 \end{array}} \right.
\end{equation}
\begin{equation}
\label{eq13} \left\{ {\begin{array}{l}
 ( {a_1^{(i)} +a_{12}^{(i)} ( {P_2^{(i)} } )^2+a_{13}^{(i)}
( {P_3^{(i)} } )^2} )P_1^{(i)} +\\ \\[1mm]
+a_{11}^{(i)}( {P_1^{(i)} } )^3+f_1^{(i)} -g_{44}^{(i)}
\frac{\partial ^2}{\partial
\,z^2}P_1^{(i)} =0\,, \\ \\[1mm]
\Biggl( \alpha _{S1}^{(i)} P_1^{(i)} +\gamma _1^{(i-1)} (
{P_1^{(i)} -P_1^{(i-1)} } )-\\
-g_{44}^{(i)} \frac{dP_1^{(i)} }{dz}
\Biggr) \Biggr|_{z=z_{i-1} } =0, \\
 \Biggl( \alpha _{S1}^{(i)} P_1^{(i)} +\gamma _1^{(i)} (
{P_1^{(i)} -P_1^{(i+1)} } )+\\
+g_{44}^{(i)} \frac{dP_1^{(i)} }{dz}
\Biggr) \Biggr|_{z=z_i } =0\,. \\
 \end{array}} \right.
\end{equation}
\end{subequations}
Equations (7) are valid for every layer designated by the index
\textquotedblleft$(i)$\textquotedblright. The electric field
$E_{3}^{(i)}=E_{3}^{d(i)}+E_{3}^{e}$. The extrapolation length
$\lambda_{i}={g_{1i}/\alpha}$ was introduced, and its experimental values fall within the
interval $\lambda=0.5\div50$\textrm{~nm }\cite{Jia:2007}.

The function
\[
f_{i}=a_{111}P_{i}^{5}+a_{123}P_{i}P_{j}^{2}P_{k}^{2}+\]
\[+a_{112}\left( {P_{i}^{3}\left( {P_{j}^{2}+P_{k}^{2}}\right)
+{P_{i}\left( {P_{j}^{4}+P_{k}^{4}}\right) }\mathord{\left/
{\vphantom {{P_i \left( {P_j^4 +P_k^4 } \right)} 2}} \right.
\kern-\nulldelimiterspace}2}\right) , \] where $i,j,k=1,2,3$ and
$i\neq j\neq k$, depends on the fifth power of the polarization and,
therefore, can be omitted for ferroelectrics with a second-kind phase
transition. It is evident that the interlayer
coupling is possible by means of the depolarization field and the
surface interaction.

In the case of multilayer periodic systems with identical
coefficients in the free energy expansion ($\gamma
_{j}^{(i-1)}=\gamma _{j}^{(i)}\equiv \gamma _{j}$), one may expect
that the polarizations inside layers with identical structures would
also be identical. Bearing in mind this simplification, we can
transform Eqs.~(7) into the following systems of equations with the
corresponding boundary conditions:
\begin{subequations}
\begin{equation}
\label{eq14} \left\{ {\begin{array}{l}
 ( {a_3^{(1)} +2a_{13}^{(1)} ( {P_i^{(1)} } )^2}
)P_3^{(1)} +a_{33}^{(1)} ( {P_3^{(1)} } )^3+\\ [2mm]+f_3^{(1)}
-g_{11}^{(1)} \frac{\partial ^2}{\partial \,z^2}P_3^{(1)} =E_3^{(1)}
\,, \\ [2mm]
 \left. {P_3^{(1)} } \right|_{z=-h_{`1} } =\left. {P_3^{(1)} } \right|_{z=0},\\
\left. {\left( {\gamma _3 ( {P_3^{(1)} -P_3^{(\ref{eq4})} }
)+g_{11}^{(1)} \frac{dP_3^{(1)} }{dz}} \right)} \right|_{z=0} =0\,. \\
 \end{array}} \right.
\end{equation}
\begin{equation}
\label{eq15} \left\{ {\begin{array}{l}
 ( {a_1^{(1)} +a_{12}^{(1)} ( {P_i^{(1)} } )^2+a_{13}^{(1)}
( {P_3^{(1)} } )^2} )P_i^{(1)} +\\ [2mm] +a_{11}^{(1)} ( {P_i^{(1)}
} )^3+f_2^{(1)} -g_{44}^{(1)} \frac{\partial ^2}{\partial
\,z^2}P_i^{(1)} =0\,, \\ [2mm]
 \left. {P_i^{(1)} } \right|_{z=-h_{`1} } =\left. {P_i^{(1)} } \right|_{z=0},\\
  \left. {\left( {\gamma _i ( {P_i^{(1)} -P_i^{(\ref{eq4})}
})+g_{44}^{(1)} \frac{dP_i^{(1)} }{dz}} \right)} \right|_{z=0} =0\,. \\
 \end{array}} \right.
\end{equation}
\end{subequations}
\begin{subequations}
\begin{equation}
\label{eq16} \left\{ {\begin{array}{l}
 ( {a_3^{(\ref{eq4})} +2a_{13}^{(\ref{eq4})} ( {P_i^{(\ref{eq4})} } )^2}
)P_3^{(\ref{eq4})} +a_{33}^{(\ref{eq4})} ( {P_3^{(\ref{eq4})} }
)^3+\\ [2mm] +f_3^{(\ref{eq4})} -g_{11}^{(\ref{eq4})} \frac{\partial
^2}{\partial \,z^2}P_3^{(\ref{eq4})} =E_3^{(\ref{eq4})} \,, \\ [2mm]
 \left. {P_3^{(\ref{eq4})} } \right|_{z=0} =\left. {P_3^{(\ref{eq4})} } \right|_{z=h_2 },\\
 \left. {\left( {\gamma _3 ( {P_3^{(\ref{eq4})} -P_3^{(1)} }
)-g_{11}^{(\ref{eq4})} \frac{dP_3^{(\ref{eq4})} }{dz}} \right)} \right|_{z=0} =0. \\
 \end{array}} \right.
\end{equation}
\begin{equation}
\label{eq17} \left\{ {\begin{array}{l}
 ( {a_1^{(\ref{eq4})} +a_{12}^{(\ref{eq4})}( {P_i^{(\ref{eq4})} } )^2+a_{13}^{(\ref{eq4})}
( {P_3^{(\ref{eq4})} } )^2} )P_i^{(\ref{eq4})} +\\ [2mm]
+a_{11}^{(\ref{eq4})} ( {P_i^{(\ref{eq4})} } )^3+f_2^{(\ref{eq4})}
-g_{44}^{(\ref{eq4})} \frac{\partial ^2}{\partial
\,z^2}P_i^{(\ref{eq4})} =0\,, \\ [2mm]
 \left. {P_i^{(\ref{eq4})} } \right|_{z=0} =\left. {P_i^{(\ref{eq4})} } \right|_{z=h_{`1} }
,\\
 \left. {\left( {\gamma _i ( {P_i^{(\ref{eq4})} -P_i^{(1)}
})+g_{44}^{(\ref{eq4})} \frac{dP_i^{(\ref{eq4})} }{dz}} \right)} \right|_{z=0} =0\,. \\
 \end{array}} \right.
\end{equation}
\end{subequations}
Equations (8) and (9) correspond to layers (1) and (2),
respectively, of the film. The polarization components for those
layers are as follows: the components $P_{3}^{(1)}$ and
$P_{3}^{(2)}$ are not oriented in the film plane, whereas the
components $( {P_{i}^{(1)}}) ^{2}=( {P_{1}^{(1)}}) ^{2}=(
{P_{2}^{(1)}}) ^{2}$ and $( {P_{i}^{(\ref{eq4})}}) ^{2}=(
{P_{1}^{(\ref{eq4})}}) ^{2}=( {P_{2}^{(\ref{eq4})}})^{2} $ do.

\begin{figure}
\includegraphics[width=7.5cm]{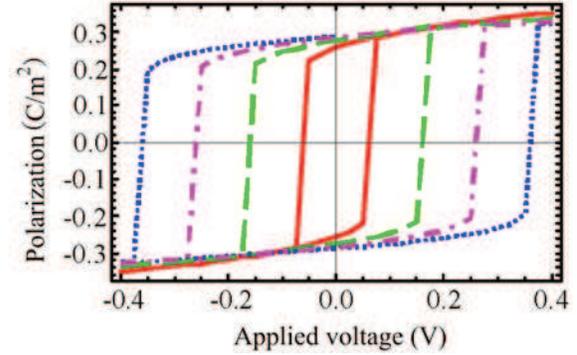}
\vskip-3mm\caption{Ferroelectric hysteresis loops in multilayer
STO/BTO films with the thickness of BTO layer $h_{\mathrm{BTO}}=5$
(solid), 9 (dashed), 13 (dah-dotted), and 17 (dotted curve) times
the lattice constant. The STO layer thickness
$h_{\mathrm{STO}}=1$~lattice constant, the interface parameter
$\gamma_{3}=\gamma_{i}=10^{5}~\mathrm{C}^{-2}\mathrm{m}^{3}\mathrm{N}$,
the misfit strain $u_{m}=-0.01$, the room temperature. Material
parameters correspond to those of the multilayer BTO/STO film  }
\end{figure}

\begin{figure}
\includegraphics[width=\column]{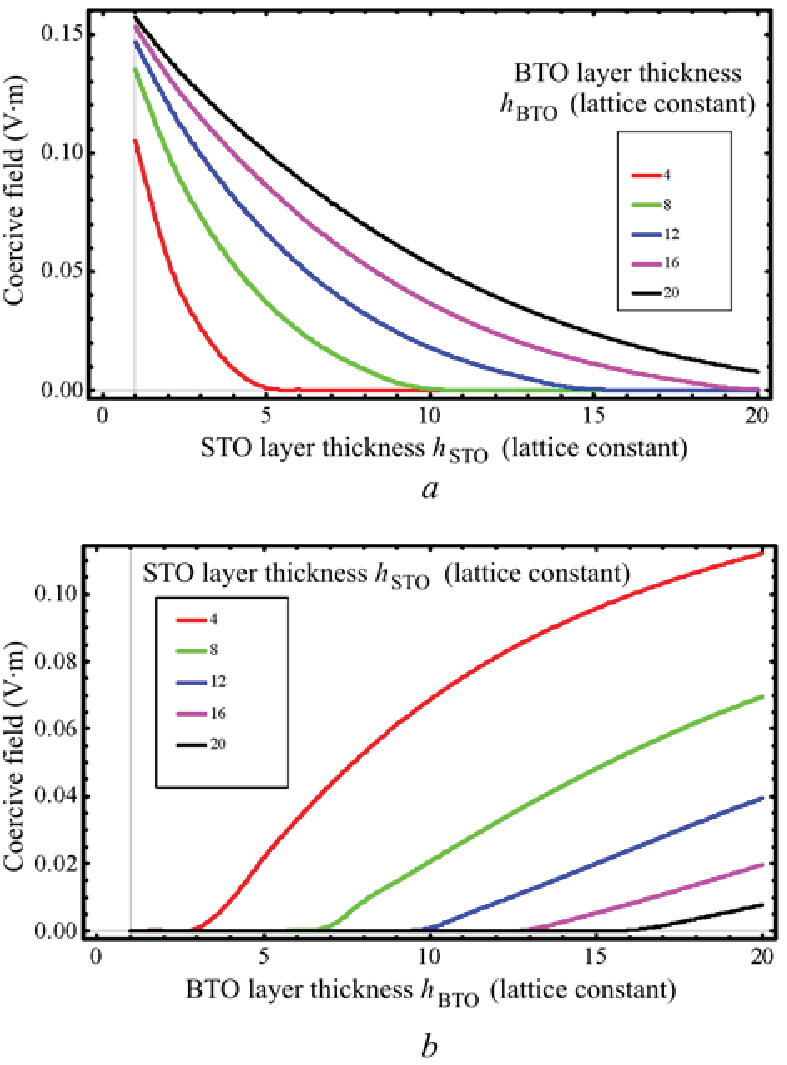}
\vskip-3mm\caption{({\it a})~Dependences of the coercive field on the
thickness $h_{\mathrm{STO}}$ of virtual ferroelectric STO layer for
various thicknesses $h_{\mathrm{BTO}}$ of a ferroelectric BTO layer.
({\it b})~Dependences of the coercive field on the thickness
$h_{\mathrm{BTO}}$ of a ferroelectric BTO layer for various
thicknesses $h_{\mathrm{STO}}$ of the virtual ferroelectric STO layer.
Both parameters are expressed in units of lattice constant.
The interface parameter
$\gamma_{3}=\gamma_{i}=10^{5}~\mathrm{C}^{-2}\mathrm{m}^{3}\mathrm{N}$,
the misfit strain $u_{m}=-0.01$, the room temperature. Material
parameters correspond to those of the multilayer BTO/STO film  }
\end{figure}

\begin{figure}
\includegraphics[width=\column]{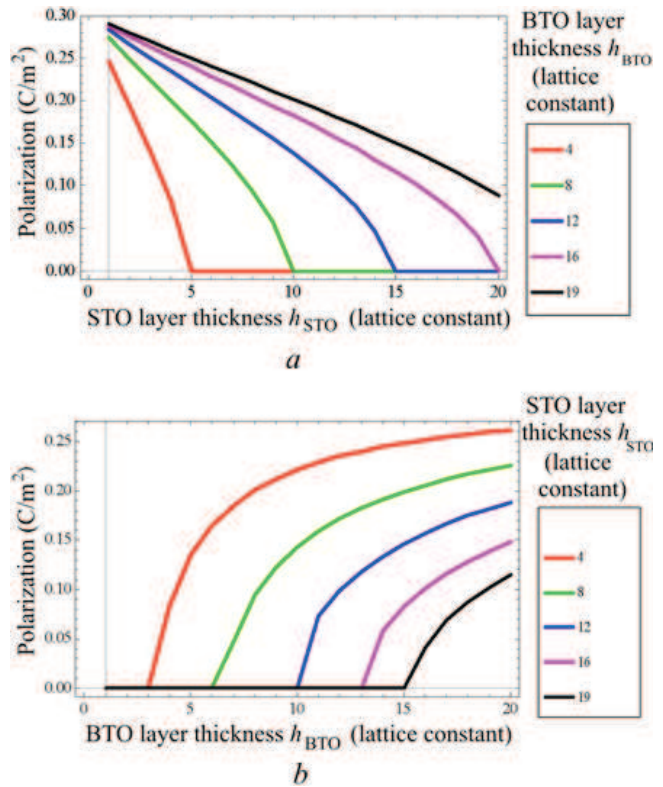}
\vskip-3mm\caption{ ({\it a})~Dependences of the spontaneous
polarization on the thickness $h_{\mathrm{STO}}$ of the virtual
ferroelectric STO layer for various thicknesses $h_{\mathrm{BTO}}$
of the ferroelectric BTO layer. ({\it b})~Dependences of the spontaneous
polarization on the thickness $h_{\mathrm{BTO}}$ of the ferroelectric
BTO layer for various thicknesses $h_{\mathrm{STO}}$ of the virtual
ferroelectric STO layer. Both parameters are expressed in units of
lattice constant. The interface parameter
$\gamma_{3}=\gamma_{i}=10^{5}~\mathrm{C}^{-2}\mathrm{m}^{3}\mathrm{N}$,
the misfit strain $u_{m}=-0.01$, the room temperature. Material
parameters correspond to those of the multilayer BTO/STO film }
\end{figure}

\section{Results of Numerical Simulation in 1D Monodomain Case}

Let us numerically examine the one-dimensional monodomain case where
$( {P_{i}^{(1)}}) ^{2}\equiv 0$ and $P_{3}^{(1,2)}\neq 0$, as is
illustrated in Fig.~1,$a$. The material parameters for a multilayer
BTO/STO film, which were used at the simulation, are quoted in
Appendix.

The most important characteristic of the polarization switching in a
multilayer film is the coercive field and the spontaneous
polarization (see an example in Fig.~2). The corresponding
dependences of the coercive field on the thickness of the virtual
ferroelectric (STO) film calculated for various thicknesses of
the ferroelectric (BTO) layer are depicted in Fig.~3,{\it a}. It is
evident that the coercive field falls down as the thickness of STO
film increases, so that STO plays a role of a dielectric layer: the
thicker the layer, the stronger is the depolarization field
(\ref{eq8}), which, in its turn, depresses the ferroelectric
properties of the BTO film. On the other hand, the coercive field
increases with the BTO film thickness (cf. different
curves).

The dependences of the coercive field on the BTO
ferroelectric film thickness for various thicknesses of the STO layer are
exhibited in Fig.~3,{\it b}. One can see that the coercive field
monotonously grows with the BTO film thickness, which
means that the thicker BTO layer possesses stronger ferroelectric
properties. On the other hand, the coercive field diminishes, when
the STO film thickness increases, and, as a result, the
depolarization field grows (cf. different curves).

The corresponding dependences of the spontaneous polarization on the
thickness of the STO virtual ferroelectric layer are shown in
Fig.~4,{\it a} for various thicknesses of the BTO film. One can see that
the spontaneous polarization monotonously decreases with the
increase of the STO film thickness, so that STO plays the role of
a dielectric layer: the thicker the layer, the stronger is the
depolarization field, which, in its turn, reduces the spontaneous
polarization of the BTO film. On the other hand, the spontaneous
polarization grows with the BTO film thickness (cf.
different curves).

The dependences of the spontaneous polarization on the BTO
ferroelectric layer thickness for various thicknesses of the STO film are shown in
Fig.~4,$b$. One can see that the spontaneous polarization
monotonously increases with the BTO film thickness and
saturates tending to a bulk value of 0.26~C/m$^{2}$ inherent to the
thinnest STO film. Therefore, the thicker the BTO layer, the
stronger are its ferroelectric properties. On the other hand, the
spontaneous polarization decreases with the increase of the STO film
thickness, because the depolarization field grows (cf. different
curves).

\section{Calculation of Multilayer-Film Phase Diagram}

To calculate the phase boundaries analytically, the linear
dielectric susceptibility must be determined. The singularity in
this parameter corresponds to the boundary between the para- and
ferroelectric BTO phases. The linear dielectric susceptibility is
determined from the linearized solution of Eq.~(9) as
$\chi_{ij}^{(i)}= {\frac{dP_{i}^{(i)}}{dE_{j}^{e}}}\Big\vert
_{E_{j}^{e}\rightarrow0}$. The results obtained are presented below.

\subsection{Boundary between ferroelectric aa-phase and paraelectric phase}

The polarization components in the aa-phase are $P_{3}^{(i)}=0$ and
$P_{1}^{(i)}=P_{2}^{(i)}\neq 0$. The boundary between the aa- and
paraelectric phases in the $\left\{ {u_{m}^{(i)},h_{1},h_{2},T}\right\}
$-coordinates is determined from the equations
\begin{subequations}
\[
0\!=\!
 \left(\! {\alpha _{S1}^{(1)} \!+\!\frac{g_{44}^{(1)} }{l_c^{(1)} }\tanh
 \left(\!
{\frac{h_1 }{2l_c^{(1)} }}\! \right)}\! \right)\!\left(\! {\alpha
_{S1}^{(\ref{eq4})}\! +\!\frac{g_{44}^{(\ref{eq4})}
}{l_c^{(\ref{eq4})} }\tanh \left(\! {\frac{h_2 }{2l_c^{(\ref{eq4})}
}} \!\right)}\! \right)\!+
\]\vspace*{-7mm}
\begin{equation}
\label{eq18}  +\!
 \gamma _1 \left(\! {\alpha _{S1}^{(1)} \!+\!\frac{g_{44}^{(1)} }{l_c^{(1)} }\tanh
\left(\! {\frac{h_1 }{2l_c^{(1)} }} \!\right)\!+\!\alpha
_{S1}^{(\ref{eq4})} \!+\!\frac{g_{44}^{(\ref{eq4})}
}{l_c^{(\ref{eq4})} }\tanh \left(\! {\frac{h_2 }{2l_c^{(\ref{eq4})}
}}\! \right)}\! \right).
\end{equation}
Here, the transverse correlation length
$l_{c}^{(i)}=\sqrt{g_{44}^{(i)}/a_{1}^{(i)}}$ is introduced, and it
is supposed that $a_{1,2}^{(1)}>0$. In the case where
$a_{1}^{(1)}>0$ and $a_{1}^{(2)}<0$, it is necessary to redefine the
correlation radius,
$\tilde{l}_{c}^{(2)}=\sqrt{-g_{44}^{(2)}/a_{1}^{(2)}}$, and rewrite
Eq.~(\ref{eq18}) as follows:
\[ 0=\left( {\alpha _{S1}^{(\ref{eq4})} -\frac{g_{44}^{(\ref{eq4})}
}{\tilde {l}_c^{(\ref{eq4})} }\tan \left( {\frac{h_2 }{2\tilde
{l}_c^{(\ref{eq4})} }} \right)} \right)\times\]
\begin{equation}
\label{eq19} \times\Bigg( {1+\frac{\gamma _1 }{\alpha _{S1}^{(1)}
+\frac{g_{44}^{(1)} }{l_c^{(1)} }\tanh \left( {\frac{h_1
}{2l_c^{(1)} }} \right)}} \Bigg)+\gamma _1\,. \tag{10b}
\end{equation}
\end{subequations}
In accordance with Eq.~(3), Eqs.~(10) depend on the misfit strain
$u_{m}^{(i)}$ and the temperature $T$ through the parameter
$a_{1}^{(i)}$. We also put $\gamma _{1}^{(i-1)}=\gamma
_{1}^{(i)}\equiv \gamma _{1}$. Note that the coefficients in
Eqs.~(10) depend not only on the misfit strain $u_{m}^{(i)}$ and the
temperature $T$, but also on the layer thicknesses $h_{i}$. In
fact, those equations determine the law of divergence for the
denominator in the susceptibility $\chi _{11}^{(2)}=
{\frac{dP_{1}^{(2)}}{dE_{1}^{e}}}\Big\vert _{E_{1}^{e}\rightarrow
0}$.

\subsection{Boundary between the ferroelectric c-phase and the paraelectric phase}

In the c-phase, the polarization components are $P_{3}^{(i)}\neq 0$
and $P_{1}^{(i)}=P_{2}^{(i)}=0$. The boundary between the c-phase
and the paraelectric phase in the $\left\{
{u_{m}^{(i)},h_{1},h_{2},T}\right\} $-coordinates is determined from
the equations
\begin{widetext}\vspace*{-7mm}
\begin{equation}
\left( {\begin{array}{l}
A_{^{{(1)}}}\varepsilon _{^{{(1)2}}}\varepsilon
_{^{{(2)}}}a_{3}^{(2)}h_{2}+A_{^{{(2)}}}\varepsilon _{^{{(2)2}}}\varepsilon
_{^{{(1)}}}a_{3}^{(1)}h_{1}+ \\[2mm]
+\displaystyle\frac{2g_{11}^{(1)}\varepsilon
_{^{{(2)2}}}A_{^{{(2)}}}\left( {\gamma _{3}\left( {\alpha
_{S3}^{(2)}L_{c}^{(2)}+g_{11}^{(2)}}\right) +\alpha
_{S3}^{(1)}\left( {g_{11}^{(2)}+L_{c}^{(2)}\left( {\alpha
_{S3}^{(2)}+\gamma _{3}}\right) }\right) }\right) -2\gamma
_{3}g_{11}^{(1)}g_{11}^{(2)}\varepsilon _{^{{(1)}}}\varepsilon
_{^{{(2)}}}A_{^{{(1)}}}}{A_{^{{(1)}}}\left( {\alpha
_{S3}^{(1)}L_{c}^{(1)}+g_{11}^{(1)}}\right) \left( {\alpha
_{S3}^{(2)}L_{c}^{(2)}+g_{11}^{(2)}}\right) +\gamma _{3}\left(
{L_{c}^{(1)}L_{c}^{(2)}\left( {\alpha _{S3}^{(1)}+\alpha
_{S3}^{(2)}}\right)
+L_{c}^{(2)}g_{11}^{(1)}+L_{c}^{(1)}g_{11}^{(2)}}\right) }+ \\[7mm]
+\displaystyle\frac{2g_{11}^{(2)}\varepsilon
_{^{{(1)2}}}A_{^{{(1)}}}\left( {\gamma _{3}\left( {\alpha
_{S3}^{(1)}L_{c}^{(1)}+g_{11}^{(1)}}\right) +\alpha
_{S3}^{(2)}\left( {g_{11}^{(1)}+L_{c}^{(1)}\left( {\alpha
_{S3}^{(1)}+\gamma _{3}}\right) }\right) }\right) -2\gamma
_{3}g_{11}^{(1)}g_{11}^{(2)}\varepsilon _{^{{(1)}}}\varepsilon
_{^{{(2)}}}A_{^{{(2)}}}}{A_{^{{(2)}}}\left( {\alpha
_{S3}^{(1)}L_{c}^{(1)}+g_{11}^{(1)}}\right) \left( {\alpha
_{S3}^{(2)}L_{c}^{(2)}+g_{11}^{(2)}}\right) +\gamma _{3}\left(
{L_{c}^{(1)}L_{c}^{(2)}\left( {\alpha _{S3}^{(1)}+\alpha
_{S3}^{(2)}}\right)
+L_{c}^{(2)}g_{11}^{(1)}+L_{c}^{(1)}g_{11}^{(2)}}\right) } \\
\end{array}}\right) =0,  \label{eq20}
\end{equation}%
\end{widetext}
where $\varepsilon _{^{{(i)}}}=\varepsilon _{0}\varepsilon
_{b}^{(i)}$, $A_{^{{(i)}}}=a_{3}^{(i)}+\frac{1}{\varepsilon
_{^{{(i)}}}}\approx \frac{1}{\varepsilon _{^{{(i)}}}}$, the
longitudinal correlation length
$L_{c}^{(i)}=\sqrt{{g_{11}^{(i)}/A}^{(i)}}$ ($i=1,2$) is introduced,
and the equalities $\gamma _{3}^{(i-1)}=\gamma _{3}^{(i)}\equiv
\gamma _{3}$ are adopted. In fact, Eqs.~(\ref{eq20}) determine the
divergence character of the denominator in the susceptibility $\chi
_{33}^{(2)}= {\frac{dP_{3}^{(2)}}{dE_{3}^{e}}}\Big\vert
_{E_{1}^{e}\rightarrow 0}$.

In the approximation
$A_{^{{(i)}}}\approx\frac{1}{\varepsilon_{^{{(i)}}}}$ obtained from
Eq.~(\ref{eq20}), we arrive at
\begin{widetext}\vspace*{-7mm}
\begin{equation}
\label{eq21}
\begin{array}{l}
 a_3^{(\ref{eq4})} h_2 +a_3^{(1)} h_1 + \\[2mm]
 +2\displaystyle\frac{\left( {\begin{array}{l}
 g_{11}^{(1)} \left( {\alpha _{S3}^{(1)} \left( {L_c^{(\ref{eq4})} \alpha
_{S3}^{(\ref{eq4})} +g_{11}^{(\ref{eq4})} } \right)+\gamma _3
L_c^{(\ref{eq4})} \left( {\alpha
_{S3}^{(\ref{eq4})} +\alpha _{S3}^{(1)} } \right)} \right)+ \\
 +g_{11}^{(\ref{eq4})} \left( {\alpha _{S3}^{(\ref{eq4})} \left( {L_c^{(1)} \alpha
_{S3}^{(1)} +g_{11}^{(1)} } \right)+\gamma _3 L_c^{(1)} \left(
{\alpha _{S3}^{(\ref{eq4})} +\alpha _{S3}^{(1)} } \right)}
\right)+\gamma _3 g_{11}^{(1)} g_{11}^{(\ref{eq4})} \left(
{\varepsilon _0 \varepsilon _b^{(1)} a_3^{(1)}
-\varepsilon _0 \varepsilon _b^{(\ref{eq4})} a_3^{(\ref{eq4})} } \right)^2 \\
 \end{array}} \right)}{\left( {\alpha _{S3}^{(1)} L_c^{(1)} +g_{11}^{(1)} }
\right)\left( {\alpha _{S3}^{(\ref{eq4})} L_c^{(\ref{eq4})}
+g_{11}^{(\ref{eq4})} } \right)+\gamma _3 \left( {L_c^{(1)}
L_c^{(\ref{eq4})} \left( {\alpha _{S3}^{(1)} +\alpha
_{S3}^{(\ref{eq4})} } \right)+L_c^{(\ref{eq4})} g_{11}^{(1)}
+L_c^{(1)} g_{11}^{(\ref{eq4})} }
\right)}\approx 0 . \\
 \end{array}
\end{equation}
\end{widetext}

\begin{figure}
\includegraphics[width=\column]{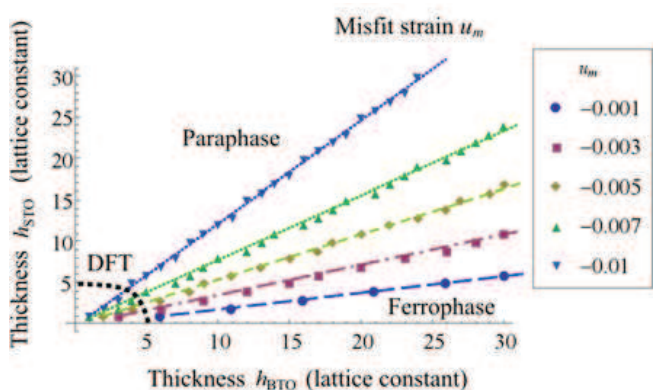}
\caption{Phase diagrams of a multilayer BTO/STO structure in the
coordinates $\left\{ {h_{\mathrm{BTO}},h_{\mathrm{STO}}}\right\} $
at room temperature and various misfit strains $u_{m}$.
$\gamma=10^{5}~\mathrm{C}^{-2}\mathrm{m}^{3}\mathrm{N}$ and
$\alpha_{Sj}^{(i)}=0$. Every straight line is a corresponding
boundary between the para- (above) and ferroelectric (below) phases
}\vskip6mm
\end{figure}

In Fig.~5, the phase diagrams of multilayer structures, which can be either
in the ferroelectric (polar) phase or in the paraelectric (nonpolar) one, are shown
as functions of the thicknesses of their STO and BTO films.\looseness=1 

The dashed line bounds the region, where the thicknesses of both films are
smaller than 5~lattice constants. Here, the coefficients of the free energy
expansion in the framework of a phenomenological theory are unknown, because
it cannot be applied owing to quantum-size effects. On such scales, the
methods of density functional theory have to be used in calculations.

The boundaries between the BTO ferroelectric phases in the $\left\{
{u_{m}^{(i)},h_{1},h_{2},T}\right\} $-coordinates were calculated
numerically.

\subsection{Phase diagram of multilayer SrTiO\boldmath$_{3}$/BaTiO$_{3}$ films}

The phase diagrams for a thin BTO film and multilayer SrTiO$_{3}$/BaTiO$_{3}$
films are shown in Fig.~6. As one can see, the boundaries between the
paraelectric (PE) and ferroelectric (FE) phases are straight lines.

It turned out that there can exist six thermodynamically stable
phases in multilayered SrTiO$_{3}$/BaTiO$_{3}$ films--one
paraelectric phase and five ferroelectric ones: a tetragonal (FEc),
two monoclinic (FEaac and FEac), and two orthorhombic (FEa and FEaa)
phases. Such a situation takes place in contrast to bulk
BaTiO$_{3}$, for which only four phases (paraelectric, tetragonal,
orthorhombic, and rhombohedral) exist \cite{bibref:1}.\looseness=1

Qualitatively, the phase diagrams of thin BTO and multilayer
SrTiO$_{3}$/BaTiO$_{3}$ films are similar. The quantitative
difference consists in that the range of the paraelectric phase
extends with the increase of the SrTiO$_{3}$ layer thickness,
because SrTiO$_{3}$ plays the role of a dielectric layer: the
thicker the layer, the stronger is the depolarization field, which,
in its turn, reduces the spontaneous polarization of the BaTiO$_{3}$
film. Simultaneously, the regions of the FEc ($P_{3}^{(i)}\neq 0$
and $P_{1}^{(i)}=P_{2}^{(i)}=0$) and FEaac ($P_{3}^{(i)}\neq 0$ and
$P_{1}^{(i)}=P_{2}^{(i)}\neq 0$) phases decrease, being governed by
the same mechanism. The phase boundaries change their shapes, as the
thickness of SrTiO$_{3}$ layers increases. It is evident that the
existence range for the PE phase is minimal for a thin film,
whereas, for multilayer films, it is larger for thicker STO layers.
STO plays the role of a dielectric interlayer, which weakens the
ferroelectric properties of a multilayer film. The boundary of the
PE phase with other phases looks like a straight line. The total
area of FE regions, where $P_{3}^{(i)}\neq 0$, decreases with the
increase of the STO layer thickness. In other words, it is
beneficial for multilayer films with thick STO layers to be
polarized in the film plane. Every phase diagram depicted in Fig.~6
includes six thermodynamically stable BaTiO$_{3}$
phases.\looseness=1

\begin{figure*}
\includegraphics[width=12.0cm]{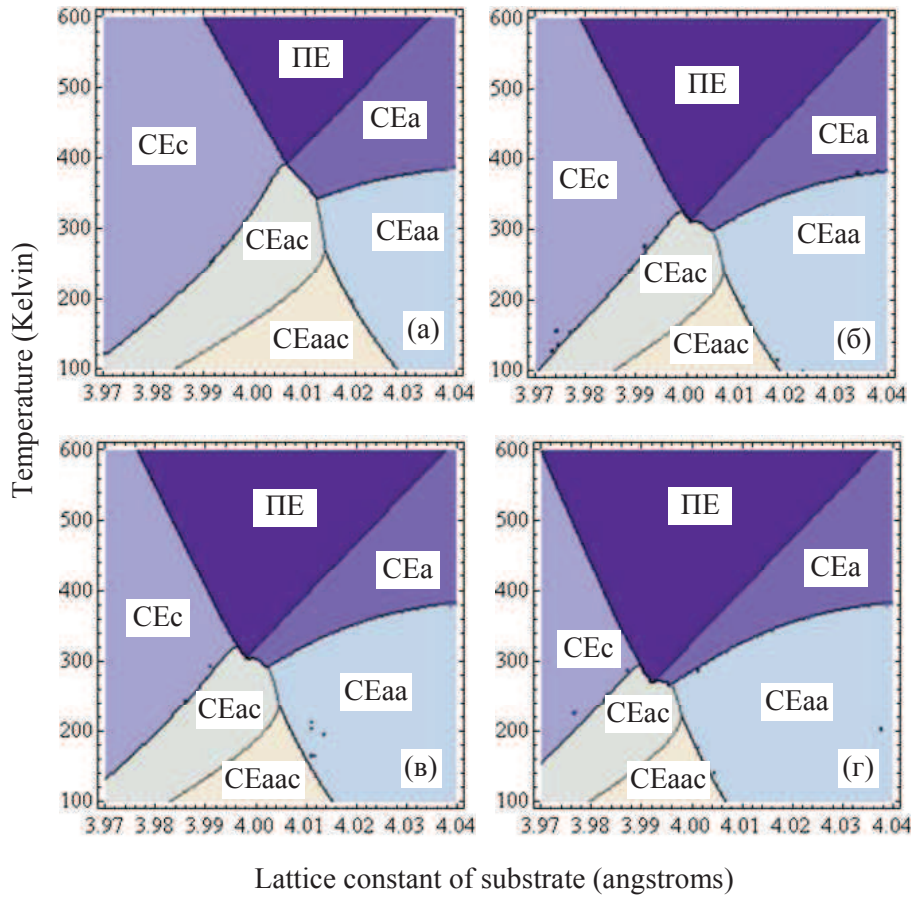}
\vskip-3mm\caption{Phase diagrams in the coordinates
\textquotedblleft substrate lattice constant versus the
temperature\textquotedblright\ for ({\it a})~thin BTO film and ({\it
b--d})~multilayer films consisting of 8-nm BTO layers and STO layers
0.8 ({\it b}), 1.2 ({\it c}), and 2.4~nm ({\it d}) in thickness (the
layers repeat periodically). The parameters of the phases are as
follows. PE phase: $P_{3}^{(i)}=P_{1}^{(i)}=P_{2}^{(i)}=0$; FEc
phase: $P_{3}^{(i)}\neq0$ and $P_{1}^{(i)}=P_{2}^{(i)}=0$; FEa
phase: $P_{3}^{(i)}=P_{2}^{(i)}=0$ and $P_{1}^{(i)}\neq 0$; FEaa
phase: $P_{3}^{(i)}=0$ and $P_{1}^{(i)}=P_{2}^{(i)}\neq0$; FEac
phase: $P_{3}^{(i)}\neq0$, $P_{1}^{(i)}=0$, and $P_{2}^{(i)}\neq0$;
and FEaac phase: $P_{3}^{(i)}\neq0$ and
$P_{1}^{(i)}=P_{2}^{(i)}\neq0$. The interface parameter
$\gamma_{3}=\gamma_{i}=10~\mathrm{C}^{-2}\mathrm{m}^{3}\mathrm{N}$
and $\alpha_{Sj}^{(i)}=0$. The parameter values were taken from work
\cite{Wang:1}  }\vskip1mm
\end{figure*}

\section{Conclusions}

The main polar properties of hysteresis loops (shape, coercive
field, and spontaneous polarization) for thin multilayer films have been
calculated. It was shown that, in the system concerned, there exists a
strong dependence of their polar properties on the thickness of SrTiO and
BaTiO layers. In particular,

\noindent1)~the coercive field monotonously grows with the BTO
film thickness; on the other hand, it decreases as the thickness of
the SrTiO$_{3}$ film grows, and, as a result, the depolarization electric field
increases;

\noindent 2)~the spontaneous polarization monotonously decreases as the
SrTiO$_{3}$ film thickness grows, so that SrTiO$_{3}$ plays the role of
a dielectric layer: the thicker the layer, the stronger is the
depolarization field, which, in its turn, reduces the spontaneous
polarization of the BaTiO$_{3}$ film; on the other hand, the spontaneous
polarization grows with the BaTiO$_{3}$ film thickness;

\noindent3)~the spontaneous polarization firstly monotonically
increases with the BTO film thickness and then saturates, by
approaching a bulk value of 0.26~C/m$^{2}$ inherent to a thin
SrTiO$_{3}$ film; the thicker the BaTiO$_{3}$ layer, the stronger
are its ferroelectric properties; on the other hand, the spontaneous
polarization decreases with the growth of the SrTiO$_{3}$ film
thickness, owing to the growth of the depolarization
field;

\noindent4)~the thickness of the BTO layer, at which the system transforms into
the ferroelectric phase, linearly grows with the STO layer
thickness;

\noindent5)~it turned out that six thermodynamically stable
BaTiO$_{3}$ phases -- one paraelectric and five ferroelectric
(tetragonal FEc, two monoclinic FEaac and FEac, and two orthorhombic
FEa and FEaa) -- can exist in multilayer SrTiO$_{3}$/BaTiO$_{3}$
films, in contrast to bulk BaTiO$_{3}$, where only four phases exist
(paraelectric, tetragonal, orthorhombic, and rhombohedral
ones).

\paragraph*{APPENDIX}


\begin{table}[!h]
\noindent\caption{Free energy parameters for bulk ferroelectric
BaTiO\boldmath$_{3}$}\vskip3mm\tabcolsep1.2pt
\noindent{\footnotesize\begin{tabular}{c c c c }
 \hline \multicolumn{1}{c}
{\rule{0pt}{9pt}Landau parameters} & \multicolumn{1}{|c}{Value}&
\multicolumn{1}{|c}{Parameters for}&
\multicolumn{1}{|c}{Value}\\%
\multicolumn{1}{c}{for free energy}&
\multicolumn{1}{|c}{}&\multicolumn{1}{|c}{elastic energy}&
\multicolumn{1}{|c}{}\\%
\hline%
$\alpha _{1}$ (10$^{5}$C$^{-2}\cdot $m$^{2}\cdot $N)&
3.61($T$--391)& $s_{11}$(C$^{-12}\cdot $m$^{2}$/N)&
8.3 \\%
$\alpha _{11}$(10$^{9}$ C$^{ -4}\cdot $m$^{6}\cdot $N)& --1.83&
$s_{12}$(C$^{-12}\cdot $m$^{2}$/N)&
--2.7 \\%
$\alpha _{12}$(10$^{9}$ C$^{ -4}\cdot $m$^{6}\cdot $N)& --2.24&
$s_{44}$(C$^{-12}\cdot $m$^{2}$/N)&
9.24 \\%
$\alpha _{111}$(10$^{10}$ C$^{ -6}\cdot $m$^{10}\cdot $N)& 1.39&
Parameters for&
Value \\
&& gradient energy&\\%
$\alpha _{112}$(10$^{9}$ C$^{ -6}\cdot $m$^{10}\cdot $N)& --2.2&
$g_{11}$(10$^{-10}$C$^{-2}\cdot $m$^{4}\cdot $N)&
2.0 \\%
$\alpha _{123}$(10$^{10}$ C$^{ -6}\cdot $m$^{10}\cdot $N)& 5.51&
$g_{44}$(10$^{-10}$C$^{-2}\cdot $m$^{4}\cdot $N)&
1.0 \\%
$Q_{11}$(C$^{ -2}\cdot $m$^{4})$& 0.11& & \\%

$Q_{12}$(C$^{ -2}\cdot $m$^{4})$& --0.043&
$G_{110}$(10$^{-10}$C$^{-2}\cdot $m$^{4}\cdot $N)&
1.73 \\%
$Q_{44}$(C$^{ -2}\cdot $m$^{4})$&
0.059&& \\%
\hline
\end{tabular}}
\end{table}

\begin{table}[h]
 \noindent\caption{Free energy parameters for bulk
virtual ferroelectric SrTiO\boldmath$_{3}$}\vskip3mm\tabcolsep1.4pt
\noindent{\footnotesize\begin{tabular}{l c c c }
 \hline \multicolumn{1}{c}
{\rule{0pt}{9pt}Parameter} & \multicolumn{1}{|c}{Measurement}&
\multicolumn{2}{|c}{Virtual ferroelectric}\\%
\cline{3-4} \multicolumn{1}{c}{}& \multicolumn{1}{|c}{
units}&\multicolumn{1}{|c}{SrTiO$_3$}&
\multicolumn{1}{|c}{Reference}\\%
\hline%
Background &&\\%
permittivity $\varepsilon _{b}$& dimensionless&3--43&[24]\\ [1.5mm]%
Coefficient $\alpha_{T}$ in&&\\ LGD expansion& 10$^{6}$ m/(F$\cdot
$C)&
1.66&[25, 26]\\ [1.5mm]%
Curie temperature&&\\
 $T_{0}$& K&36 &[25, 26]\\ [1.5mm]%
Temperature of&&\\
quantum-mechanical&&\\
vibrations $T_{q}$& K&100&[25, 26]\\ [1.5mm]%
LGD gradient&&\\
coefficient $g$&10$^{-10}$ В$\cdot $m$^{3}$/C&
1--5&[27]\\ [1.5mm]%
Coefficients $\alpha_{ij}$&10$^{9}$
m$^{5}$/(C$^{2}\cdot $F)&$\alpha _{11}=8.1$&\\%
in LGD expansion&& $\alpha_{12}=2.4$ &[25, 26]\\ [1.5mm]%
Coefficients $\alpha _{ijk}$&&\\
in LGD expansion &10$^{12}$
m$^{9}$/(C$^{4}\cdot $F)&$\sim 1$&Undefined \\ [1.5mm]%
Electrostriction&
m$^{4}$/C$^{2}$&$Q_{11}=0.051$\\%
coefficients $Q_{ijkl}$&&$Q_{12}=~-0.016$&\\%
&&$Q_{44}=0.020$&[28]\\ [1.5mm]%
 Elastic rigidity&10$^{11}$ N/m$^{2}$&$c_{11}=3.36$\\%
  $c_{ij}$ &&$c_{12}=1.07$ &\\%
 &&$c_{44}=1.27$ &[28]\\ [1.5mm]%
Elastic compliance&10$^{-12}$ m$^{2}$/N&$s_{11}=3,89$\\%
$s_{ij}$&&$s_{12}=-1.06$ &\\%
&& $s_{44}=8.20$&[28]\\%
 \hline
\end{tabular}}
\end{table}

\rezume{%
ПОЛЯРНІ ВЛАСТИВОСТІ ТА ПЕТЛІ ГІСТЕРЕЗИСУ\\ У БАГАТОШАРОВИХ ТОНКИХ
ПЛІВКАХ ТИПУ \\СЕГНЕТОЕЛЕКТРИК/ВІРТУАЛЬНИЙ \\СЕГНЕТОЕЛЕКТРИК}{Є.А.
Єлісєєв, М.Д. Глинчук, Г.М. Морозовська,\\ Я.В. Яковенко}
{Використовуючи феноменологічну теорію Ландау--Гінзбурга--Девоншира,
розраховано вплив деформацій невідповідності, поверхневої енергії та
розмірних ефектів на фазові діаграми, полярні властивості та петлі
гістерезису у багатошарових тонких плівках типу
сегнетоелектрик/віртуальний сегнетоелект\-рик. Вперше досліджено
вплив пружних деформацій, що виникають на межі тонка
плівка--підкладка внаслідок невідповідності сталих ґратки плівки та
підкладки, на фазові діаграми багатошарових тонких плівок складу
віртуальний сегнетоелектрик SrTiO$_{3}$/сегнетоелект\-рик
BaTiO$_{3}$. Виявилося, що у багатошарових плівках складу
SrTiO$_{3}$/BaTiO$_{3}$ можуть існувати шість термодинамічно стійких
фаз BaTiO$_{3}$ (параелектрична, тетрагональна FEc, дві моноклінні:
FEaac та FEac, дві орторомбічні: FEa та FEaa сегнетоелектричні фази)
на відміну від об'ємного BaTiO$_{3}$, де існують лише чотири фази
(кубічна, тетрагональна, орторомбічна та ромбоедрична). Розраховано
основні полярні властивості петель гістерезису (форма, коерцитивне
поле і спонтанна поляризація) у тонких багатошарових плівках
SrTiO$_{3}$/BaTiO$_{3}$. Показано, що у системі існує сильна
залежність полярних властивостей від товщини шарів SrTiO$_{3}$ і
BaTiO$_{3}$ та пружних деформацій невідповідності, причому
SrTiO$_{3}$ відіграє роль діелектричного прошарку: чим товщий
прошарок, тим сильніше поле деполяризації,\linebreak яке, у свою
чергу, зменшує спонтанну поляризацію плівки BaTiO$_{3}$.}


\begin{thebibliography}{99}
\bibitem{27}The order of magnitude
$g\sim10^{-10}~\mathrm{V\cdot m}^{3}/\mathrm{K}$ corresponds to
ordinary \textquotedblleft rigid\textquotedblright\ ferroelectrics
like PTO, whereas $g\sim 10^{-9}~\mathrm{V\cdot m}^{3}/\mathrm{K}$
seems to be more typical of virtual ferroelectrics, ``soft''
ferroelectrics, and ferroelectric semiconductors.\vskip1.7mm
\end{thebibliography}
\end{document}